\documentstyle[12pt,epsfig]{article}
\newlength{\dinwidth}
\newlength{\dinmargin}
\newcommand{\ba}{\begin{array}}
\newcommand{\ea}{\end{array}}
\newcommand{\be}{\begin{equation}}
\newcommand{\ee}{\end{equation}}
\newcommand{\bea}{\begin{eqnarray}}
\newcommand{\eea}{\end{eqnarray}}
\newcommand{\gsim}{\mathrel{\mathop{\kern 0pt \rlap
  {\raise.2ex\hbox{$>$}}} \lower.9ex\hbox{\kern-.190em $\sim$}}}

\def\ben{\begin{equation}}
\def\een{\end{equation}}
\def\bea{\begin{eqnarray}}
\def\eea{\end{eqnarray}}
\def\nn{\nonumber}

{\rm

\begin{document}

\thispagestyle{empty}
\addtocounter{page}{-1}
\vskip-0.35cm
\begin{flushright} 
{\tt hep-th/yymmxxx}
\end{flushright}
\vspace*{0.2cm}
\centerline{\Large \bf Non-trivial 2d space-times from matrices}\footnote{Based
on talk at {\em ``International Workshop on String Theory''},
Khajuraho, December 2004.}

\vspace*{1.0cm} \centerline{\bf Sumit R. Das}
\vspace*{0.7cm}
\centerline{\it Department of Physics and Astronomy,}
\vspace*{0.2cm}
\centerline{\it University of Kentucky, Lexington, KY 40506 \rm USA}
\vspace{0.7cm}
\centerline{\tt das@pa.uky.edu}

\vspace*{0.8cm}
\centerline{\bf Abstract}
\vspace*{0.3cm}

Solutions of matrix quantum mechanics have been shown to describe time
dependent backgrounds in the holographically dual two dimensional
closed string theory. We review some recent work dealing with
non-trivial space-times which arise in this fashion and discuss
aspects of physical phenomena in them.

\vspace*{0.5cm}

\newpage

%%%%%%%%%%%%%%%%%%%%%%%%%%%%%%%%%%%%%%%%%%%%%%%%%%%%%%%%%%%%%%%%%%%%%%%%%%%
\section{Introduction}
%%%%%%%%%%%%%%%%%%%%%%%%%%%%%%%%%%%%%%%%%%%%%%%%%%%%%%%%%%%%%%%%%%%%%%%%%%%

One of the most profound lessons of recent research in String Theory
is that gravitational physics typically has a holographic
description \cite{'tHooft:1993gx}
in terms of lower dimensional theories which do not
contain gravity. This is a manifestation of the duality between closed
strings and open strings and the most celebrated example is AdS/CFT
correspondence \cite{Maldacena:1997re}
where the presence of horizons result in a truncation
of the open string theory to its low energy field theory limit. 

The earliest example of such a holographic correspondence is two
dimensional noncritical string theory \cite{ceqone} . Here the
holographic model is matrix quantum mechanics which has no space. The
closed string theory lives in one space and one time dimension - the
space dimension arising from the space of eigenvalues of the matrix in
the the large-N limit \cite{Das:1990ka}. Recently it has been realized
that this is also an example of open-closed duality
\cite{McGreevy:2003kb}.  Two dimensional string theory is a toy model
of string theory - but a useful toy model because of its
solvability. In particular, the emergence of space can be understood
in an explicit fashion using the techniques of collective field theory
\cite{Jevicki:1979mb, Das:1990ka}.

As is well known, the ground state of the quantum mechanical system is
dual to the static linear dilaton vacuum of the closed string theory
with a liouville wall provided by an additional tachyon background. In
addition to the ground state, time-dependent backgrounds have been
known for a while \cite{Minic:1991rk}. Recently such solutions have
been proposed as models of cosmology \cite{Karczmarek:2003pv}.  Using
standard techniques, the dual space-times may be constructed which
provide interesting models for posing and answering questions about
time-dependent backgrounds in string theory. In this talk I will
review some results in this area.

Some of these models can be used as useful tools to understand issues
of particle production in time dependent backgrounds
\cite{Karczmarek:2004ph, Das:2004hw} and display characteristic
stringy effects \cite{Das:2004hw}.  Another class of solutions
naturally lead to space-like boundaries. Normally these space-times
would be regarded as geodesically incomplete. However, the underlying
fundamental description of dynamics in terms of the matrix model
precludes any extension of the space-times beyond these boundaries.
The results show that the construction of extra dimensions from the
degrees of freedom of a holographic model can be rather non-trivial.

\section{Space-time around the ground state}

Let us first recapitulate how space-time emerges around the ground
state of the model.

The dynamical variable of the model is a single $N \times N$ matrix
$M_{ij}(t)$ 
and there is a constraint which restricts the states to be singlets.
In the singlet sector and in the double scaling limit \cite{Gross:1990ay}
matrix quantum
mechanics reduces to a theory of an infinite number of 
fermions with the single particle
hamiltonian given by
\ben
H = {1\over 2}[p^2 - x^2]
\label{one}
\een
where we have adopted conventions in which the string scale
$\alpha^\prime = 1$ for the bosonic theory and $\alpha^\prime =
{1\over 2}$ in the fermionic theory. The fermi energy in this rescaled
problem will be denoted by $-\mu$.

In the classical limit, the system is equivalent to an incompressible
fermi fluid in phase space. The ground state is the static fermi
profile
\ben
(x-p)(x+p)=2\mu
\label{two}
\een

The dual closed string theory is best obtained by rewriting the theory
in terms of the collective field $\phi (x,t)$ which is defined as the
density of eigenvalues of the original matrix. 
\ben
\partial_x \phi (x,t) = {1\over N}{\rm Tr}\delta (M(t)-x\cdot I)
\een
At the classical level
the action of the collective field is given by
\ben
S = N^2\int dxdt~[{1\over 2}{(\partial_t \phi)^2 \over (\partial_x \phi)}
  - {\pi^2 \over 6} (\partial_x \phi)^3 - (\mu - {1\over
    2}x^2)\partial_x \phi]
\label{aone}
\een
This is of course a theory in $1+1$ dimension, the spatial dimension
arising out of the space of eigenvalues.

All solutions represented by Fermi seas do not necessarily
appear as {\em classical} solutions to collective field
theory\cite{Dhar:1992cs}
. However fermi surfaces with {\em quadratic profiles}
do. Nonquadratic profiles can be still represented as states of the
quantum collective theory, albeit with large fluctuations which
survive the classical limit \cite{Das:1995gd}.
The ground state profile (\ref{two}) is such a quadratic profile and
the classsical solution is
\ben
\partial_x \phi_0 = {1\over \pi}
	{\sqrt{x^2-2\mu}}~~~~~\partial_t\phi_0 = 0 
\label{atwo}
\een

The space-time which is generated may be obtained 
by looking at the dynamics of fluctuations of the collective field
around the classical solution. For later use, we will in fact expand
around an {\em arbitrary} classical solution $\phi_0 (x,t)$
\ben
\phi(x,t) = \phi_0 (x,t) + {1\over N}\eta (x,t)
\label{asix}
\een
The action for these fluctuations at the quadratic level may be
written as
\ben
S^{(2)}_\eta = {1\over 2} \int dt dx~{\sqrt{g}}g^{\mu\nu}\partial_\mu
\eta \partial_\nu \eta
\label{aseven}
\een
where $\mu,\nu = t,x$. The line element determined by 
$g_{\mu\nu}$ is conformal to 
\ben
ds^2 = -dt^2 + {(dx + {\partial_t \phi_0 \over \partial_x \phi_0}dt)^2 
\over (\pi \partial_x \phi_0)^2}
\label{aeight}
\een
Therefore, {\em regardless of the classical solution} the spectrum is
always a single massless scalar in one space dimension given by $x$.
The metric can be determined only upto a conformal factor. However, as
we will see below, the global properties of the space-time can be
determined from the nature of the classical solution.

The classical 
interaction hamiltonian is purely cubic when expressed in terms of the
fluctuation field $\eta$ and its canonically conjugate momentum
$\Pi_\eta$, 
\ben
H^{(3)}_\eta = \int dx [{1\over 2}\Pi_\eta^2 \partial_x\eta + {\pi^2
    \over 6} (\partial_x\eta)^3]
\label{anine}
\een

Around the ground state, the metric (\ref{aeight}) is given by
\ben
ds^2 = -dt^2 + {dx^2 \over x^2 -2\mu}
\label{aten}
\een
The perturbative fluctuations live in the region $|x| > {\sqrt{2\mu}}$
and the field $\eta$ satisfies Dirichlet boundary condition at the
``mirrors'' given by $x = \pm {\sqrt{2\mu}}$. The fields on the
``left'' and ``right'' side are decoupled. The physics of these fields
is made transparent by choosing Minkowskian coordinates
$(\sigma,\tau)$ which in this case are
\ben
t = \tau ~~~~~~~~~ x = \pm {\sqrt{2\mu}} \cosh \sigma
\label{aeleven}
\een
In these coordinates 
\ben
ds^2 = -d\tau^2 + d\sigma^2
\een

\begin{figure}[ht]
\centerline{\epsfxsize=1.5in
\epsfysize=2.0in
   {\epsffile{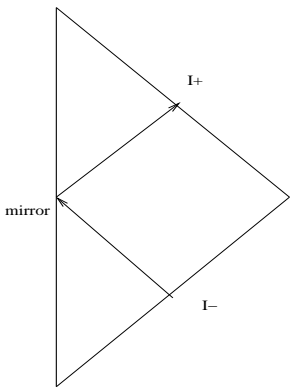}}
}
\caption{Penrose diagram of space-time produced by ground state
  solution showing an incoming ray getting reflected at the mirror}

\label{penrose_ground}
\end{figure}

The field $\eta$ may be now thought of being made of {\em two} fields,
$\eta_{S,A}(x,t)$ each of which live in the region $x > 0$
\ben
\eta_{S,A} (x,t) = {1\over 2}[\eta (x,t) \pm \eta(-x,t)]
\label{atwelve}
\een
In terms of the Minkowskian coordinates, solutions to the
linearized equations are plane waves $\eta_{S,A} \sim
e^{-i\omega(t\pm\sigma)} \eta_{S,A} (\omega)$ and these fourier
components are related to the two spacetime fields - the tachyon $T$
and the axion $C$ - which appear in the
standard formulation of Type 0B string theory 
\cite{Takayanagi:2003sm} :
\bea
T(\omega) & = & (\pi/2)^{-i\omega/8}~{\Gamma(i\omega/2) \over
\Gamma(-i\omega/2)}~\eta_S (\omega) \nn \\
C(\omega) & = & (\pi/2)^{-i\omega/8}~{\Gamma((1+i\omega)/2) \over
\Gamma((1-i\omega)/2)}~\eta_A (\omega)
\label{athirteen}
\eea
Recall that we are working in string units. These transforms therefore
imply that the position space fields are related by a transform which
is non-local at the string scale. Therefore points on the Penrose diagram should be thought of as smeared over the string scale. But then, this should be true of any Penrose diagram drawn in a string theory.

In any case, the space-time generated is quite simple. The Penrose
diagram is that of two dimensional Minkowski space with a mirror at
$\sigma = 0$, as shown in Figure (\ref{penrose_ground}). 
The fluctuations are massless particles which come in
from ${\cal I}^-_{L,R}$, get reflected at the mirror and arrive at
${\cal I}^+_{L,R}$. In terms of the Minkowskian coordinates the
interaction hamiltonian becomes
\ben
H_3 = \int d\sigma{1\over 2\sinh^2\sigma}[{1\over
    2}{\tilde{\Pi_\eta}}^2\partial_\sigma\eta + {\pi^2 \over
    6}(\partial_\sigma\eta)^3] 
\label{bone}
\een 
The interactions therefore vanish at $\sigma = \infty$ and
are strong at $\sigma = 0$ - this gives rise to a non-trivial wall
S-matrix. 

In the Type 0B string theory interpretation, the Penrose diagram has
to be folded across the center and a point which is localised in
$\sigma$ space is smeared out in the string theory space over string
length. 

As emphasized above, the metric is determined only upto a conformal transformation. So long as the conformal transformation is non-singular, this is sufficient to draw Penrose diagrams. A conformal transformation would, however, mix up the space $\sigma$  and time $\tau$ and it would appear that this leads to an ambiguity. The special property of the space and time coordinates defined above is that the {\em interaction Hamiltonian is time-independent} with this choice and a conformal transformation would destroy this property. This makes the physics transparent and easy to compare with string theory results. Of course a different coordinatization with a time dependent Hamiltonian is physically equivalent and should be compared with the string theory results in an appropriately chosen gauge.

\section{Moving Fermi Seas}

One class of time-dependent solutions are
generated from (\ref{two}) by the action of $W_\infty$ symmetries in
phase space \cite{Das:2004hw}. 
In the classical limit, these are represented by moving
fermi seas :
\ben
x^2-p^2 + \lambda_-~e^{-rt}(x+p)^r + \lambda_+ ~e^{rt} (x-p)^r +
\lambda_+\lambda_- (x^2-p^2)^{r-1} = 2\mu~.
\label{three}
\een
where $r$ is a non-negative integer and 

$\lambda_\pm$ are finite parameters. The $r=1$ solution was 
considered in \cite{Karczmarek:2003pv} and proposed as a model of cosmology.
Formally, the state of the fermion system 
is related to the
ground state $|\mu\rangle$ by 
\ben
|\lambda\rangle = {\rm exp}[i\lambda Q]|\mu\rangle ~,
\een
where $Q$ denotes the $W_\infty$ charge which generates this solution.
However this state is not normalizable and therefore not contained in the
Hilbert space of the model. Rather, this corresponds to
a deformation of the hamiltonian of the theory to 
\ben
{H}^\prime = e^{-i\lambda Q}{H}e^{i\lambda Q} ~.
\label{vthree}
\een

In the following we will study
the dual spacetimes which arise from the solutions with $r=1$ and
$r=2$. For these solutions the fermi surfaces are quadratic, so that

these correspond to classical solutions of collective field theory. 

\begin{figure}[ht]
\centerline{\epsfxsize=3.0in
\epsfysize=3.0in
   {\epsffile{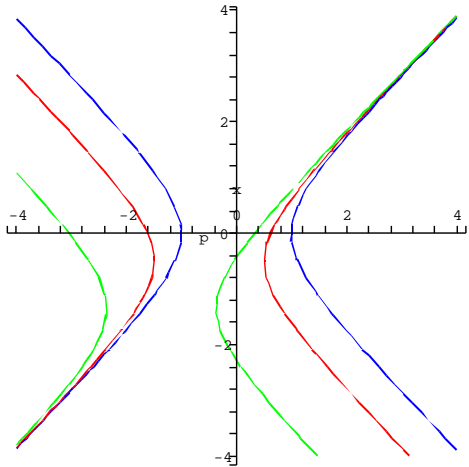}}
}
\caption{The draining/flooding sea solution. The fermi surface moves
  to the left and downwards. Fermions which are in the $x < 0$ region
  at $t = -\infty$ are drained out while fermions originally in the $x
  > 0$ flood in, eventually crossing over to the other side.}
\label{r1fermi}
\end{figure}

For $r=1$ and $\lambda_-=0, \lambda_- > 0$ 
we can choose the origin of time to choose
$\lambda_+ = 2$. Then the classical collective field is 
\ben
\partial_x \phi_0  =  {1\over \pi}{\sqrt{(x+e^t)^2 - 2\mu}} ~~~~~~~~~
\partial_t \phi_0  =  -e^t~\partial_x \phi_0
\label{athree}
\een
We will call this the ``draining/flooding fermi sea'' solution.
The fermi surface is displayed in (\ref{r1fermi}) 

\begin{figure}[ht]
\centerline{\epsfxsize=3.0in
\epsfysize=3.0in
   {\epsffile{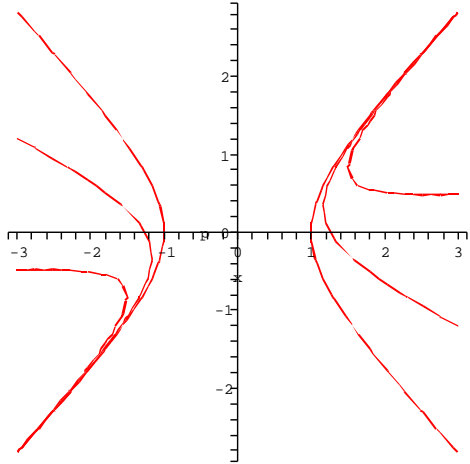}}
}
\caption{The closing hyperbola solution. At late times the hyperbola
  closes on into itself drianing all the fermions}
\label{r2fermi1}
\end{figure}

The physics of the 
$r=2$ solutions was considered in \cite{Das:2004aq}.
For $\lambda_- = 0$ with $\lambda_+ < 0$ which can be chosen
to be $-2$ we have
\ben
\partial_x \phi_0  =  {1\over \pi(1 + e^{2t})}{\sqrt{x^2-(1+e^{2t})}}
~~~~~~
\partial_t \phi_0  =  - {x e^{2t} \over 1+e^{2t}}~\partial_x \phi_0
\label{afive}
\een
Here we have rescaled $x$ and $t$ to set $2\mu = 1$.
This will be called the ``closing hyperbola'' solution shown and
explained in Figure (\ref{r2fermi1}).

\begin{figure}[ht]
\centerline{\epsfxsize=3.0in
\epsfysize=3.0in
   {\epsffile{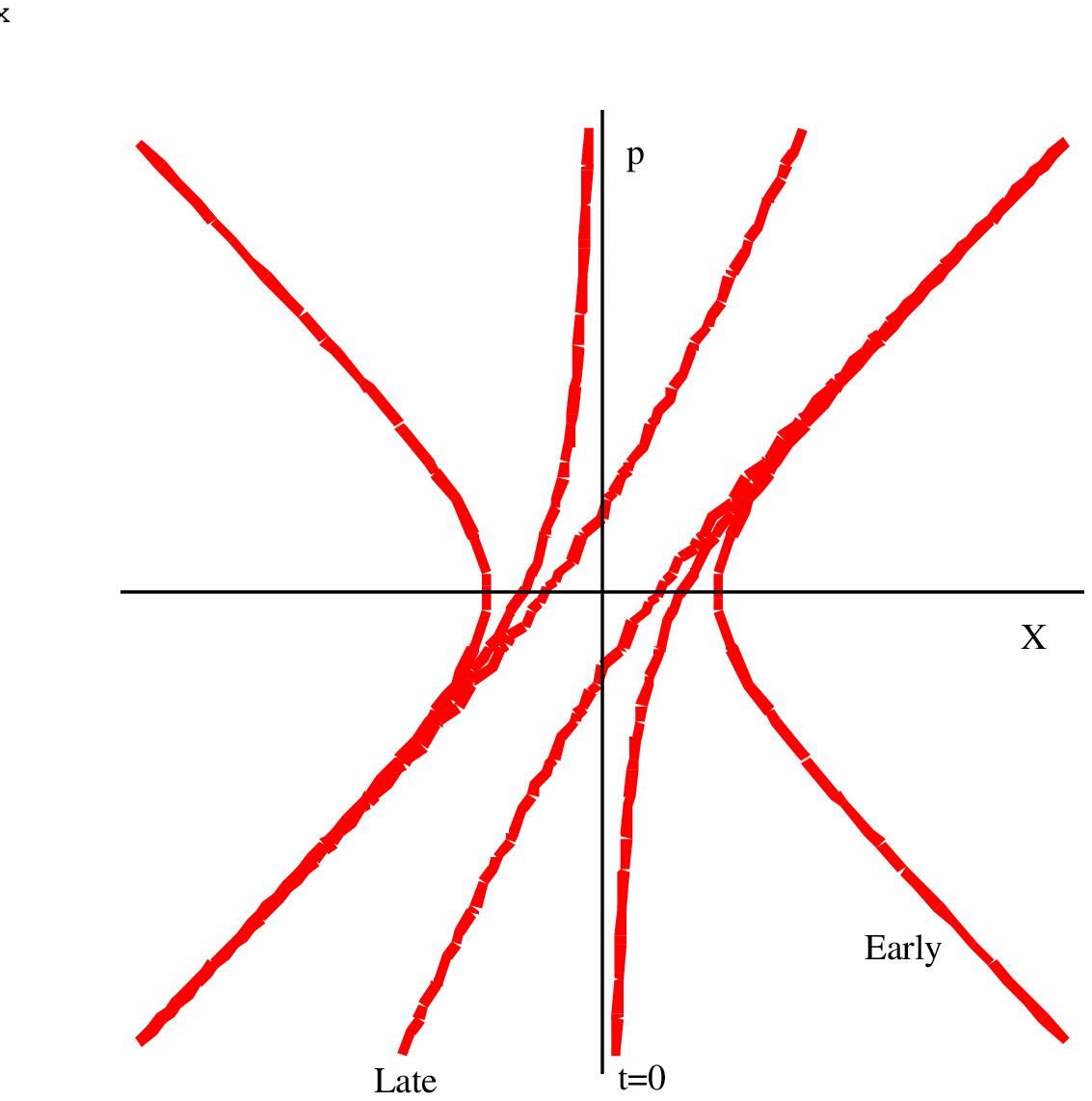}}
}
\caption{The opening hyperbola solution. At $t=0$ the surface becomes
vertical and tips over for $t > 0$ eventually asymptotic to $x=p$ line}
\label{r2fermi}
\end{figure}

Finally, for 
$r=2$ and $\lambda_- = 0$ with $\lambda_+ > 0$ which can be chosen
to be $2$ we have
\ben
\partial_x \phi_0  =  {1\over \pi(1-e^{2t})}{\sqrt{x^2-(1-e^{2t})}}
~~~~~~
\partial_t \phi_0  =  {x e^{2t} \over 1-e^{2t}}~\partial_x \phi_0
\label{afour}
\een
where we have again set $2\mu =1$.
This will be called the ``opening hyperbola'' solution. The phase
space fermi surfaces are shown in (\ref{r2fermi})

What do these solutions mean in the worldsheet string theory ? This
question has not been settled yet, but it is clear that this
corresponds to a closed string tachyon condensation. Proposals for the
worldsheet perturbations corresponding to these solutions have been
made in \cite{Karczmarek:2003pv, Karczmarek:2004ph, Das:2004aq}. 
This issue will not be discussed here further.

\section{Space-time structures}

We will study the nature of space-times generated by these solutions 
by looking at fluctuations around them. The strategy will be to pass
to Minkowskian coordinates in each case so that the Penrose diagrams
may be drawn easily. As argued above, points on these diagrams should be considered fuzzy at the string scale.

\subsection{The draining/flooding sea}

The draining/flooding sea solution is a simple modification of the
ground state, but encodes non-trivial physics. 

Let us first consider
the fluctuations around the left branch of the fermi sea. It is
trivial to see that the Minkowskian coordinates $(\sigma,\tau)$ are
given by
\ben
t = \tau~~~~~~~~x= -\cosh \sigma - e^\tau
\label{btwo}
\een
The mirror is always at $\sigma = 0$. However in the original
``space'' $x$, the mirror is {\em moving} towards the asymptotic
region. 

The moving mirror problem has served as an excellent toy model for
various properties of quantum fields in time dependent geometres,
e.g. Hawking radiation from a collapsing black hole. Here, the moving
mirror arises from within a string theory. In fact the mirror is
``soft'' since there can be tunelling across the barrier of the
inverted oscillator potential. In the following, however, we will restrict our attention to perturbative processes which ignore tunneling.

At early times, the solution is identical to the static
solution so that it is natural to define a space coordinate $q$
appropriate for incoming modes by
\ben
x = -\cosh q
\label{bthree}
\een
The Penrose diagram in $(\tau,\sigma)$ space is the same as in the
ground state. On ${\cal I}^-$, defined as $\sigma_- = 
(\sigma - \tau) \rightarrow
\infty$ with $\sigma_+ = \sigma + \tau$ finite, 
the two coordinates $q$ and $\sigma$
coincide.
However, on ${\cal I}^+$ 
\ben
\sigma \pm \tau = (q \pm t) + \log(1-2e^{-(q-t)})
\label{bfour}
\een
which shows that the mirror hits ${\cal I}^+$ at a point
$q_- = q-t = \log (2)$. 
The interaction hamiltonian is once again
given by (\ref{bone}) and therefore vanish on ${\cal I}^\pm$.

Equation (\ref{bfour}) implies a nontrivial relation between the in
modes starting on ${\cal I}^-$ 
and the out modes which end on ${\cal I}^+$ after reflection
from the mirror and results in particle creation
\cite{Karczmarek:2004ph},\cite{Das:2004hw}.
However, as
discussed below, the fact that our moving mirror is embedded in a {\em
string theory} has nontrivial consequences.

The fluctuations of the right branch of the fermi surface have a
slightly different physics. Now the mirror is {\em receding} away from
the asymptotic region. As a result modes coming in from ${\cal I}^-$
will reflect back to $x = \infty$ 
if they start off at early times, but will not be
able to catch up with the mirror if they start off at late
times. These modes then cross over to the other side of the potential.
Therefore ${\cal I}^+$ has two pieces : ${\cal I}^+ = {\cal I}^+_1 +
{\cal I}^+_2$ where ${\cal I}^+_1$ is located at large positive values
of $x$ at late times while ${\cal I}^+_2$ is located at large negative
values of $x$ at late times. In the Type 0B string theory
interpretation, however, a crossing of the barrier means a
transformation of the relative strengths of the tachyon and axion
fields. 

\subsection{The closing hyperbola solution}

For both the ground state and the moving mirror solutions, the
original matrix model time $t$ is identical to the time $\tau$ in
terms of which the metric is Minkowskian. A direct consequence of this
is that the Penrose diagrams are quite similar. In contrast, the
opening and closing hyperbola solutions involve a non-trivial
relationship between $\tau$ and $t$. The necessary coordinate
transformations may be obtained using the techniques of
\cite{Alexandrov:2003uh}. 
For the closing hyperbola
solution one has, for the right side
\ben
x = \cosh \sigma {\sqrt{1+e^{2t}}}~~~~~~~~e^\tau = {e^t \over
  {\sqrt{1+e^{2t}}}}
\label{bfive}
\een
This immediately shows that as $-\infty < t < \infty$ the time $\tau$
has the range $ -\infty < \tau < 0$. Since the dynamics of the matrix
model ends at $t = \infty$ the resulting space-time appears to be
geodesically incomplete with a space-like boundary at $\tau = 0$.

\begin{figure}[ht]
\centerline{\epsfxsize=2.0in
\epsfysize=2.0in
   {\epsffile{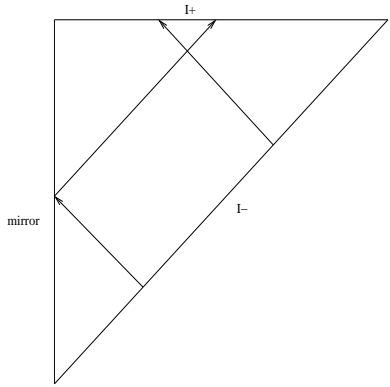}}
}
\caption{Penrose diagram for the closing hyperbola solution showing
  two classes of null rays}
\label{penrose_closing}
\end{figure}

There is always a mirror.  Fluctuations coming in from ${\cal I}^-$
along $\sigma_+ = \sigma + \tau = \tau_0$ will get reflected by the
mirror at $\sigma = 0$ so long as $\tau_0 < 0$. For $\tau_0 > 0$ this
ray cannot reach the mirror before time ends - rather it directly hits
the space-like boundary at $\tau = 0$. The Penrose diagram with these
two classes of rays is shown in Figure (\ref{penrose_closing}). 

In the original space-time $(t,x)$ however, small fluctuations of the
Fermi surface are {\em always} reflected back to large $x$
region. The exact trajectory of a point on the fermi surface is given
by
\ben
x(t) = e^t \cosh \tau_0 + {1\over 2} e^{-t} e^{\tau_0}
\label{bsix}
\een
It may be easily verified that $dx/dt = 0$ always has a solution, and
the point turns back in $x$ space {\em before it reaches the
  mirror} at $\sigma = 0$. In fact the whole of $\tau = 0$ surface is
at $x = \infty$.

The interaction hamiltonian which evolves in time $\tau$ may be
obtained by starting from the expression (\ref{anine}). The result is
\ben
H_3 = {\partial \over \partial \tau} = ({dt \over d\tau})
{\partial \over \partial t} = 
({dt \over d\tau})
\int d\sigma {1\over (\partial x / \partial \sigma)^2}[{1\over
    2}{\tilde{\Pi_\eta}}^2\partial_\sigma\eta + {\pi^2 \over
    6}(\partial_\sigma\eta)^3] 
\label{bseven}
\een
where we have used the fact that $\tau = \tau (t)$ only. In this case
this reduces exactly to equation (\ref{bone}). Thus the coupling on
the space-like boundary is exactly the same as on any other space-like
slice - strong at $\sigma =0$ and small in the asymptotic
region. In particular this implies that the coupling is strong even
when $x$ is large.
Our conclusions should be trustworthy in the region of large
$\sigma$. 

As in the case of the ground state, we have chosen the space and time such that the {\em interaction Hamiltonian is time-independent}. This property will not be maintained under a conformal transformation.

\subsection{The opening hyperbola space-time}

The space-time generated by the closing hyperbola solution is even
more non-trivial. It is clear from the profile of the fermi surface
that the ``mirror'' disappears at the time $t=0$. In fact we now
require two patches of Minkowskian space-time to describe time
evolution. Let us first discuss the space-time perceived by
fluctuations of the right branch of the fermi surface. 
For $t < 0$ the Minkowskian coordinates $(\tau,\sigma)$ are given by
\ben
x = \cosh \sigma {\sqrt{1-e^{2t}}}~~~~~~~e^\tau = {e^t \over
{\sqrt{1-e^{2t}}}}
\label{beight}
\een
The range $-\infty < t < 0$ maps into the complete range in Minkowski
time $-\infty < \tau < \infty$ and the line $\sigma + \tau = \infty$
with finite $\sigma_-$ marks a horizon of this coordinate system. For
$t > 0$ we have another patch with
\ben
x = \sinh \sigma {\sqrt{e^{2t}-1}}~~~~~~~e^{-\tau} = {e^t \over
{\sqrt{e^{2t}-1}}}
\label{bnine}
\een
Note that for $t > 0$ the entire $x$ space is covered by this
coordinate patch whereas for $t < 0$ one has to have an additional
patch which is relevant to the fluctuations of the left branch of the
fermi surface.

Once again, space-time ends on a space-like surface at $\tau = 0$
in the second patch. 
The exact trajectory of a point on the fermi
surface is given by
\ben
x(t) = -e^t \sinh \tau_0 + {1\over 2} e^{-t}e^{\tau_0}
\label{bten}
\een
In the $(\tau,\sigma)$ space {\em all} the trajectories turn around at
the mirror at $\sigma = 0$. Upon reflection, they go across the horizon
at $\sigma_+ = \infty$ of the first patch into the second patch and
end up on the $\tau = 0$ surface of the second patch. For $\tau_0 < 0$
these end up in the $x \rightarrow \infty $ 
region while for $\tau_0 > 0$ they end up in the $x \rightarrow
-\infty$ region.

The trajectories look rather different in the original $(t,x)$
space-time. 
It is straightforward to see from (\ref{bten}) 
that (i) for $-\infty < \tau_0 < -{1\over 2}\log~2$ they reflect back
in $x$ space at a negative value of $t$. (ii) For $ -{1\over 2}\log~2
< \tau_0 < 0$ they reflect back in $x$ space for positive $t$ and
(iii) For $\tau_0 > 0$ the rays never reflect back in $x$ space but
proceed to the other side. 

\begin{figure}[ht]
\centerline{\epsfxsize=3.0in
\epsfysize=3.5in
   {\epsffile{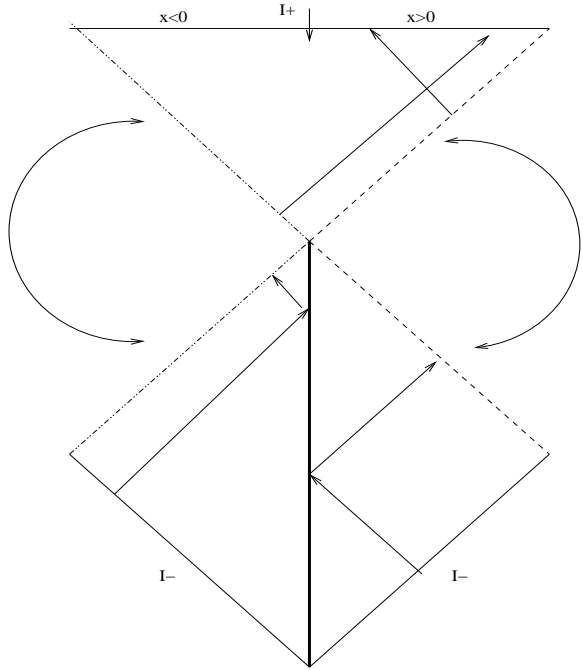}}
}
\caption{Penrose diagram for the opening hyperbola solution showing
  two classes of null rays. The identifications are indicated by
curved arrows}
\label{penrose_opening}
\end{figure}

The story for fluctuations originating on the left branch of the
hyperbola is exactly similar. Putting these together we can draw a
Penrose diagram as in Figure (\ref{penrose_opening}).

Finally, on ${\cal I}^+$  the interaction hamiltonian (\ref{bseven})
is again time-independent : 
\ben
H_3 = 
{\partial \over \partial \tau} = \int d\sigma{1\over 2\cosh^2\sigma}[{1\over
    2}{\tilde{\Pi_\eta}}^2\partial_\sigma\eta + {\pi^2 \over
    6}(\partial_\sigma\eta)^3] 
\label{beleven}
\een 
While the whole of $\tau =0$ surface has $x = \pm \infty$ the
interactions are nontrivial. However they again vanish in the
asymptotic region of large $\sigma$ and this is where our results are
trustworthy. 

\subsection{The nature of boundaries}

For the two classes of closing and opening hyperbola solutions we
concluded that the space-time have space-like boundaries. This is
based on the fact that the boundary is a $\tau = 0$ surface in terms
of the Minkowskian time. Normally one would extend the
space-time beyond this surface and complete it. However, this extension
does not make sense in this theory. Recall that space-time is a
derived concept in this model and the true dynamics is that of
infinite number of eigenvalues of the matrix evolving in time
$t$. This dynamics is interpreted as a dynamics of a field in one more
dimension. Extension of the space-time beyond the surfaces $\tau =0$
would mean that we extend the fundamental time evolution beyond $t =
\infty$ and this does not make any sense.

In two dimensions, any metric is conformally flat. Since our model
contains only massless scalars, the free part of the Hamiltonian is
not sensitive to the conformal factor. One could possibly choose a
conformal factor which is singular at the boundary and makes the
space-time geodesically complete. This should lead to an equivalent
description of the physics. The problem with this is that the
necessary conformal transformation would make the interaction
Hamiltonian time dependent and possibly singular at {\em all points in
space} at late times. In our choice the interaction Hamiltonian is
time-independent. While it does not vanish everywhere on the
space-like boundary, it does vanish in the asymptotic region at all
times. This makes the physics more transparent. For fluctuations
around the ground state, this is a useful choice since this
facilitates comparison with worldsheet results. In the case of
non-trivial solutions, such a comparison is not yet possible.

The presence of space-like boundaries in the solutions above should be
interpreted in view of these comments. The unambiguous statement is
that the boundaries are space-like in a choice of coordinates in which
the interaction Hamiltonian is time-independent.

\section{Aspects of particle production and stringy effects}

In the moving mirror background, modes receive large blue(red) shifts
as they get reflected and this results in particle production. A
standard way to study this is to consider the components of the ``out''
energy momentum tensor in the ``in'' vacuum, using the anomalous
transformation properties in two dimensions. 
In this case ${\cal I}^-$ is
parametrized by $\sigma_+=\sigma +\tau$ (as defined in equation
(\ref{btwo}) which is identical to $q_+ = q+t$ in this region. The ``in''
vacuum is defined in terms of the modes $u_{in}=e^{-i\omega (\tau \pm
\sigma)}$. On the other hand ${\cal I}^+$ is defined by $\sigma_+
\rightarrow \infty$ and parametrized by either $\sigma_-$ or $q_-$
which are related by (\ref{bfour}). This means that one measure of the
outgoing energy-momentum flux is the quantity $< T_{q_-q_-}>_{in}$
where the quantity $T_{q_-q_-}$ is defined in terms of the derivatives
of the field with respect to $q_-$ with constant large $\sigma_+$
(which is not the same as constant $q_+$). This quantity can be
calculated in terms of $<T_{\sigma_-\sigma_-}>_{in}$ by performing a
coordinate transformation from $\sigma_\pm$ to $(q_-,\sigma_+)$. The
relation (\ref{bfour}) shows that this is a conformal transformation,
so that the anomaly relation gives 
\ben 
<T_{q_-q_-}>_{in} = ({\partial
\sigma_- \over \partial q_-})^2 <T_{\sigma_-\sigma_-}>_{in} + {1\over
24\pi} \{ \sigma_-, q_- \}_S
\label{btwelve}
\een
where $\{ A, B \}_S $ denotes the Schwarzian derivative. In our case
\ben {1\over 24\pi} \{ \sigma_-, q_- \}_S = {1\over 48\pi}{4
e^{-q_-}(1-e^{-q_-}) \over (1-2e^{-q_-})^2}
\label{bthirteen}
\een
In usual quantum field theory one would set
$<T_{\sigma_-\sigma_-}>_{in} = 0$ - by definition the quantity is
normal ordered in the in vacuum.  
Therefore the
energy-momentum flux of particles is simply given by (\ref{bthirteen})

However, here we are dealing with a string theory and the quantity
under consideration is the torus worldsheet partition function. Unlike
usual quantum field theory, this quantity has a well defined meaning
since string theory does not have ultraviolet divergences in physical
quantities - and this quantity is nonvanishing and finite in
theories without target space supersymmetry. This important fact seems
to be at odds with the corresponding calculation in collective field
theory.

The answer to this apparent puzzle is well known. The action of
collective field theory given in (\ref{aone}) is only the {\em
classical} action. At the loop level there are additional terms in the
action \cite{Jevicki:1979mb},
which are calculated concretely by carefully making the change
of variables from the matrix to collective fields. The explicit term
which is relevant for our purposes is given by the singular expression
\ben 
\Delta S_{(1)} = -{1\over 2\pi}\int
dx~(\partial_x\phi)~[\partial_x\partial_{x^\prime}~\log
|x-x^\prime|]_{x=x^\prime}
\label{bfourteen}
\een
This is clearly subdominant in the large-N and hence $\hbar$ expansion
of the theory. At one loop one can substitute $\phi \rightarrow
\phi_0$. For fluctuations around the ground state this term leads to
an infinite contribution to the one loop ground state energy. The
singular term cancels the infinite contribution from integrating out
fluctuations (the usual one loop diagram in the field theory of
$\eta$) leaving a finite result which is in exact agreement with
worldsheet calculations \cite{Das:1990ka}.
This result may be expressed in terms of a
ground state energy momentum tensor
\ben
<T_{++}>_{gs} = <T_{--}>_{gs} = -{1\over 48\pi}
\label{bfifteen}
\een
This finite one loop ground state energy must be clearly taken into
account in the calculation of the energy-momentum flux from the moving
mirror \cite{Das:2004hw}. 
In this problem $\sigma_\pm$ are Minkowskian coordinates -
this means that the quantity $<T_{\sigma_-\sigma_-}>_{in}$ which
appears in the anomaly relation (\ref{btwelve}) is precisely the
ground state quantity $<T_{--}>_{gs}$ given by
(\ref{bfifteen}). Substituting this in (\ref{btwelve}) one finds the
$q_-$ dependent contribution from the Schwarzian derivative in the
second term is cancelled by the first term, leaving with {\em exactly
the ground state answer}
\ben
<T_{q_-q_-}>_{in} = -{1\over 48\pi}
\label{bsixteen}
\een
Since we continue to use $\sigma_+$ as the relevant coordinate at late
times the value of $<T_{\sigma_+\sigma_+}>$ is unchanged and is also
$-1/48\pi$ so that in this sense there is no energy flux, even though
there is particle production. What has happened is that the constant
one loop energy density in the moving coordinate system $\sigma,\tau$
translates into an energy flux in the coordinates $(q_-,\sigma_+)$
which cancels the energy flux due to particle production. This is a
very stringy effect - since the additional one loop term in the
collective field action is a reflection of the underlying string
theory.

Instead of computing the quantities considered above one could also
calculate the expectation values of canonical quantities in the
original $(t,q)$ coordinate system leading to $<T_{q_-q_-}>_{in}$
which are defined in terms of derivatives ${\partial \over
\partial q_-}|_{q_+}$. Since the transformation between $q_+$ and
$\sigma_+$ is not a canonical transformation for large $q_+$ one
cannot use the anomaly relation as above. Rather the calculation has
to be done ab initio with careful regularization. This has been done
in \cite{Mukhopadhyay:2004ff}.
The result is that while divergent terms cancel, the finite result has
a nontrivial $q_-$ dependence.

The questions of physical effects like particle production in the
other geometries considered in this talk are more complicated and have
not been fully addressed yet.  

An issue which is closely related to particle production is the
question of thermality. This requires a time-symmetric model. In
\cite{Karczmarek:2004yc} a time symmetric version of $r=1$ solution
was shown to lead to Hartle-Hawking type states, despite nontrivial
time dependence.  

\section{Other solutions}

There are several other classes of time-dependent solutions of the
$c=1$ which have been studied using methods similar to those discussed
above. One interesting class of such solutions involve finite size
droplets in phase space \cite{Ernebjerg:2004ut}. In fact there has
been recent progress in understanding the question of black hole (non)
formation in this theory and Hawking radiation \cite{Friess:2004tq}.
In a slightly different direction, \cite{Takayanagi:2004yr} have found
that a {\em time-like} linear dilaton background in two dimensional
string theory can be also described in terms of a matrix model, which
is in fact quite similar to the standard one. Some non-perturbative
aspects of time-dependence have been considered in
\cite{Alexandrov:2004cg}. 
The outstanding problem
is to discuss these backgrounds in the worldsheet
formulation. It is likely that recent work on minimal string theory
\cite{Seiberg:2003nm}
can shed some light on this problem.

\section{Acknowledgemnts} 
I would like to thank Josh Davis, Joanna Karczmarek, Finn Larsen and
Partha Mukhopadhyay for very enjoyable collaborations and numerous
discussions.  I would also like to thank Atish Dabholkar, Avinash
Dhar, Rajesh Gopakumar, Gary Horowitz, Antal Jevicki, Gautam Mandal,
Hiroshi Ooguri, Ashoke Sen and Andy Strominger for conversations and
comments during seminars at various places. Finally I thank the
organizers of this conference for a immaculately organized
meeting. This work has been partially supported by a NSF grant
PHY-0244811 and a DOE contract DE-FG-01-00ER45832.

\end{document}